\documentclass{aa}

\usepackage{graphicx}
\usepackage{units}
\usepackage{xcolor}
\usepackage{ulem}
\usepackage{txfonts}
\usepackage{hyperref}
\begin{document}

   \title{Revealing magnetic field structure at the surfaces of protoplanetary disks via near-infrared circular polarization}

   \author{I. de Langen\inst{1}
          \and
          R. Tazaki\inst{1,2}}

   \institute{Anton Pannekoek Institute for Astronomy, University of Amsterdam, Science Park 904, 1098 XH Amsterdam, The Netherlands\\
              \email{delangen@mps.mpg.de}
             \and
             Astronomical Institute, Graduate School of Science, Tohoku University, 6-3 Aramaki, Aoba-ku, Sendai 980-8578, Japan\\
             }


 
  \abstract
   {Magnetic fields play a fundamental role in the dynamical evolution of protoplanetary disks, in particular via magnetically induced disk winds. The magnetic field structure at the disk surface is crucial for driving the disk winds; however, it is still poorly understood observationally.}
   {We explore a new method to probe the magnetic field structure at the disk surface using near-infrared (NIR) circular polarization. Near-infrared circular polarization arises when unpolarized stellar light is scattered by magnetically aligned grains at the disk surface. In this study, we aim to clarify to what extent the observed circular polarization pattern can be used to diagnose the magnetic field structure. }
   {We first calculated light scattering properties of aligned spheroids, and the results were then used to create expected observational images of the degree of circular polarization at a NIR wavelength.}
   {Magnetically aligned grains can produce circular polarization,  particularly when the field configuration deviates from a purely toroidal field. We find that disk azimuthal dependence of the degree of circular polarization tends to exhibit a double peaked profile when the field structure is favorable for driving disk winds by centrifugal force. We also find that even if the disk is spatially unresolved, a net circular polarization can possibly be nonzero. We also show that the amplitude of circular polarization is strongly dependent on grain composition and axis ratio.} 
   {Our results suggest that circular polarization observations would be useful to study the magnetic field structure and dust properties at the disk surface.}

    \titlerunning{Revealing magnetic field structure at the surface of PPD via NIR CP}
    
  \maketitle
%

\section{Introduction}

Magnetic fields are a crucial factor in determining how protoplanetary disks form, evolve, and disperse. Recent studies suggest that magnetohydrodynamical (MHD) disk winds play an important role in disk evolution. The occurrence of MHD disk winds is strongly dependent on the magnetic field structure at the disk surface. A poloidal magnetic field structure with a large angle with respect to the disk normal would give rise to disk winds via centrifugal force \citep{Blandford1982}, while a more toroidal field would drive winds through a magnetic pressure gradient \citep[e.g.,][]{Bai2017}. Therefore, to study how MHD disk winds are launched, it is essential to understand the magnetic field geometry at the disk surface. 

Observing the magnetic field in protoplanetary disks is a notoriously difficult task. Polarized thermal emission from magnetically aligned grains has been the most successful technique in the interstellar medium \citep{lazarian2007,andersson2015}. However, this technique is not always reliable in protoplanetary disks, because larger grains that are responsible for thermal emission are not necessarily aligned with the magnetic field \citep{tazaki2017,Yang2021}. Alternatively, molecular line polarization might contain information about both $B$-field orientation and strength, but requires exquisite sensitivity, which can only marginally be reached with current telescopes \citep{teague2021,harrison2021,Stephens2020}. Recently, \cite{yangli2022} explored a new technique, finding that linear polarization angles of near-infrared (NIR) disk scattered light carry information on the magnetic field, as long as micron-sized grains at the disk surfaces are magnetically aligned. One of the great advantages of this technique is its sensitivity to the disk-surface magnetic field, which is crucial for understanding MHD disk winds.  

Although \cite{yangli2022} mainly focused on linearly polarized scattered light arising from magnetically aligned grains, we can also expect circular polarization (CP) of scattered light \citep[e.g.,][]{BandermannKemp1973, Fukushima2020}.  However, there have been no quantitative studies on how disk-scattered light is circularly polarized at near-IR and to what extent it can be used for diagnosing disk magnetic fields. 

In this study, we aim to trace the magnetic field structure in protoplanetary disks using circular polarized scattered light. To achieve this goal, we will create synthetic maps showing the degree of circular polarization throughout a protoplanetary disk and study how and to what extent the $B$-field is reflected in the circular polarization pattern.

This paper is organized as follows. We introduce our method and the dust model in Sect. \ref{sec:methods}. In Sect. \ref{sec:results}, we present the CP patterns. We discuss the implications for future observations in Sect. \ref{sec:discussion}. Finally, we conclude in Sect. \ref{sec:summary}. 

\section{Methods and models} \label{sec:methods}
To calculate the CP of disk-scattered light in the NIR, we follow a similar method to \citet{yangli2022}, where the authors calculated linearly polarized scattered light by aligned grains at the disk surface. In this study, we assume single scattering; the stellar radiation will be scattered by a grain at the disk surface only once before reaching the observer.
\subsection{Circular polarization calculation} \label{sec:cpmethod}
At the disk surface, dust particles scatter unpolarized stellar light to the observer. The scattered light is in general linearly and circularly polarized. The polarization of scattered light is described by the Stokes parameter $(I_{s}, Q_{s}, U_{s}, V_{s})$, which can be calculated from the Stokes parameter of incident light ($I_{i},Q_{i},U_{i},V_{i}$) and the scattering matrix \citep{Bohren1983}: 
\begin{equation} \label{eq:scat}
    \begin{pmatrix}
        I_{s} \\
        Q_{s} \\
        U_{s} \\
        V_{s} 
    \end{pmatrix}
     \propto
     \begin{pmatrix}
         S_{11} & S_{12} & S_{13} & S_{14} \\
         S_{21} & S_{22} & S_{23} & S_{24} \\
         S_{31} & S_{32} & S_{33} & S_{34} \\
         S_{41} & S_{42} & S_{43} & S_{44} 
     \end{pmatrix}
         \begin{pmatrix}
        I_{i} \\
        Q_{i} \\
        U_{i} \\
        V_{i} 
    \end{pmatrix}.
\end{equation}
The scattering matrix elements $S_{ij}$ describe how the polarization state of light changes upon scattering by a grain. Considering unpolarized incident light from the star, the degree of circular polarization in scattered light $V_{s}/I_{s}$ is equal to $S_{41}$/$S_{11}$. The scattering matrix elements are given by \citep{Bohren1983}
\begin{gather}
    S_{11} = \frac{1}{2} \, (|S_1|^{2} + |S_2|^{2}+ |S_3|^{2}+ |S_4|^{2}), \label{eq:s11}\\
    S_{41} = \textrm{Im}\{S_{2}^{\ast}S_4 + S_{3}^{\ast}S_{1}\} \label{eq:s41} ,
\end{gather}
where $S_{i} \, (i=1,2,3,4)$ represent the scattering amplitude. To calculate Eqs. \eqref{eq:s11} and \eqref{eq:s41}, we assume the dipole approximation (Rayleigh limit), where a grain has to be small compared to the wavelength. We also assume oblate dust grains with an axis ratio of 1.5. With these assumptions, we can find analytical expressions for the scattering amplitude $S_i$ \citep{Bohren1983}. In the Rayleigh limit, $S_{41}/S_{11}$ does not depend on the size parameter, which is determined by grain size and wavelength, but depends solely on dust composition (refractive index) and the axis ratio of the oblate grain. The validity of the dipole approximation is discussed in Sect. \ref{sec:large}.

We assume a dust composition model proposed by \citet{Birnstiel2018}, where each grain consists of a mixture of water ice \citep{Warren2008}, carbonaceous material, troilite \citep{Henning1996}, and astronomical silicate \citep{Draine2003}. For carbonaceous material, we consider two possibilities: one is refractory organics \citep{Henning1996}, as with \citet{Birnstiel2018}, and the other is amorphous carbon \citep{Zubko1996}. Although dust composition including refractory organics has been commonly used in the literature \citep{Pollack1996,Alessio2001,Kataoka2014,Birnstiel2018}, a composition including amorphous carbon has also been used to model disk observational quantities \citep[e.g., DIANA opacity:][]{Woitke2016}. The detailed effect of dust composition on CP is discussed in Sect. \ref{sec:other}. Using \verb|optool| \citep{Dominik2021}, we derived an effective refractive index at $\lambda = 1.6 \,\rm{\mu m}$  using the Bruggeman mixing rule \citep{Bruggeman1935}. As a result, we find a refractive index of $m = 1.5625 + i0.0204$ and $m = 2.0217 + i0.4308$ for a mixture comprising refractory organics and amorphous carbon, respectively. 

\subsection{Disk model and magnetic field geometries}
\begin{figure}
\resizebox{\hsize}{!}{\includegraphics{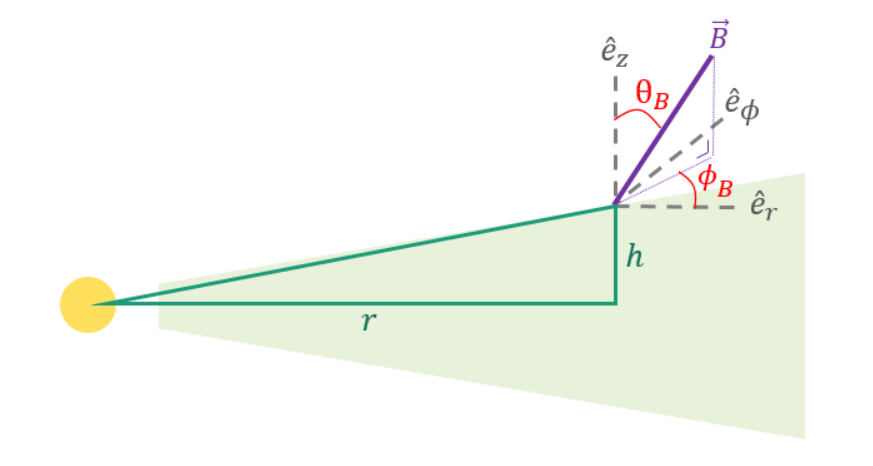}}
\caption{Schematic view of the disk model and magnetic field geometry. The location of a scattering grain at the disk surface is defined by $r$ and $h$. The magnetic field orientation is specified by $\phi_{B}$ and $\theta_{B}$, measured in the cylindrical coordinate frame $(\hat{e}_{r},\hat{e}_{\phi},\hat{e}_{z})$.}
\label{fig:disk_cartoon}
\end{figure}

To model the disk-scattering surface at NIR wavelengths, we assumed an axisymmetric flared disk. We determined the disk flaring geometry motivated by NIR observations \citep{Avenhaus2018}. 
By fitting ring structures seen in scattered light images, \citet{Avenhaus2018} determined the heights of disk scattering surfaces. As a result, they obtained an empirical fit to the ratio of the height of the scattering surface $h$ and the midplane distance from the star $r$ as follows:
\begin{equation}
    \frac{h}{r} = 0.1617 \cdot \left(\frac{r}{100~\mathrm{au}}\right)^{0.219}.
\end{equation}
At the disk surface, the oblate grains are aligned with the magnetic field. For the sake of simplicity, we assume that grains are perfectly aligned with their short axis parallel to the local magnetic field direction. We introduce two parameters ($\theta_B, \phi_B$) to specify the local magnetic field direction, as done in \citet{yangli2022} (see Fig. \ref{fig:disk_cartoon}). For example, when $\theta_B=0^\circ$, the magnetic field is vertical. When $\theta_B=90^\circ$ and $\phi_B=0^\circ$, the field is radial. When $\theta_B=90^\circ$ and $\phi_B=90^\circ$, the field is azimuthal (toroidal). By considering the range of [$0^\circ,90^\circ$] for both $\theta_B$ and $\phi_B$, we can mimic various magnetic field structures at the disk surface.

Given the disk geometry, the orientation of aligned oblate grains, and the inclination angle, we can calculate scattering angles at each location of the disk surface \citep[see e.g.,][for the case of spherical grains]{Stolker2016}. By applying these scattering angles to the CP calculation method described in Sect. \ref{sec:cpmethod}, we can make a projected disk image of the degree of CP. We also calculated the linear polarization of scattered light and confirmed that our code can successfully reproduce the results presented in \citet{yangli2022}.

\section{Results} \label{sec:results}
We calculated the degree of circular polarization produced by aligned dust grains at the disk surface for a set of assumed magnetic field structures and analyzed the imprint of the $B$-field on the CP pattern. 

\subsection{Integrated CP values} \label{sec:integrated}
\begin{figure}
\resizebox{\hsize}{!}{\includegraphics{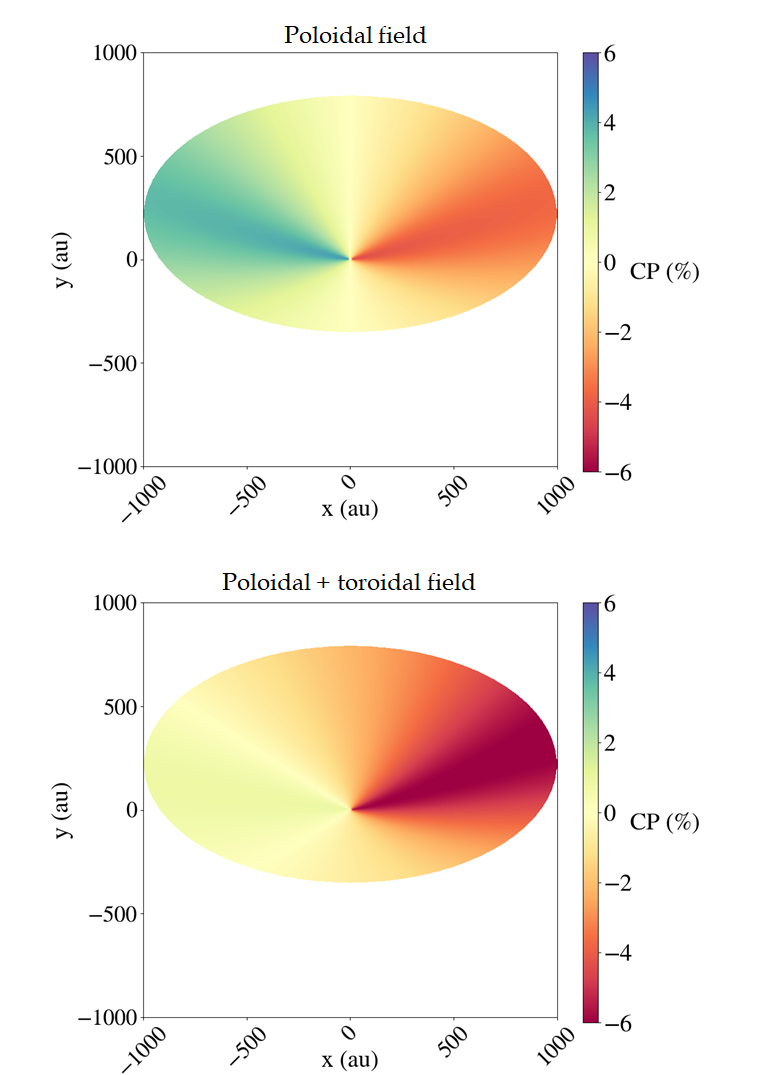}}
\caption{Synthetic maps of the CP produced at the protoplanetary disk surface, assuming a dust composition with amorphous carbon, $i=55^{\circ}$ and $\theta_{B}=45^{\circ}$. The upper panel shows the result for a poloidal field ($\phi_{B} = 0^{\circ}$), resulting in a symmetric CP pattern. The bottom panel shows the result for a magnetic field with a poloidal and toroidal component ($\phi_{B} = 60^{\circ}$), where the symmetry is broken.}
\label{fig:disks}
\end{figure}

We show two synthetic maps of the degree of circular polarization projected at the surface of the disk model in Fig. \ref{fig:disks}. For both images, we assumed an inclination of $i=55^{\circ}$, and polar magnetic field angle $\theta_{B}=45^{\circ}$, but we assumed different values of $\phi_{B}$. The upper panel represents a disk with $\phi_{B}=0^{\circ}$, which corresponds to a purely poloidal field. The CP pattern that arises from this disk surface is mirror symmetric (except for the sign) with respect to the disk minor axis. In the lower panel, we added a toroidal component to the magnetic field by increasing $\phi_{B}$ to $60^{\circ}$. In this situation, the mirror symmetry is broken, which demonstrates the effect of a different magnetic field structure on the CP pattern. 

\begin{figure}
\resizebox{\hsize}{!}{\includegraphics{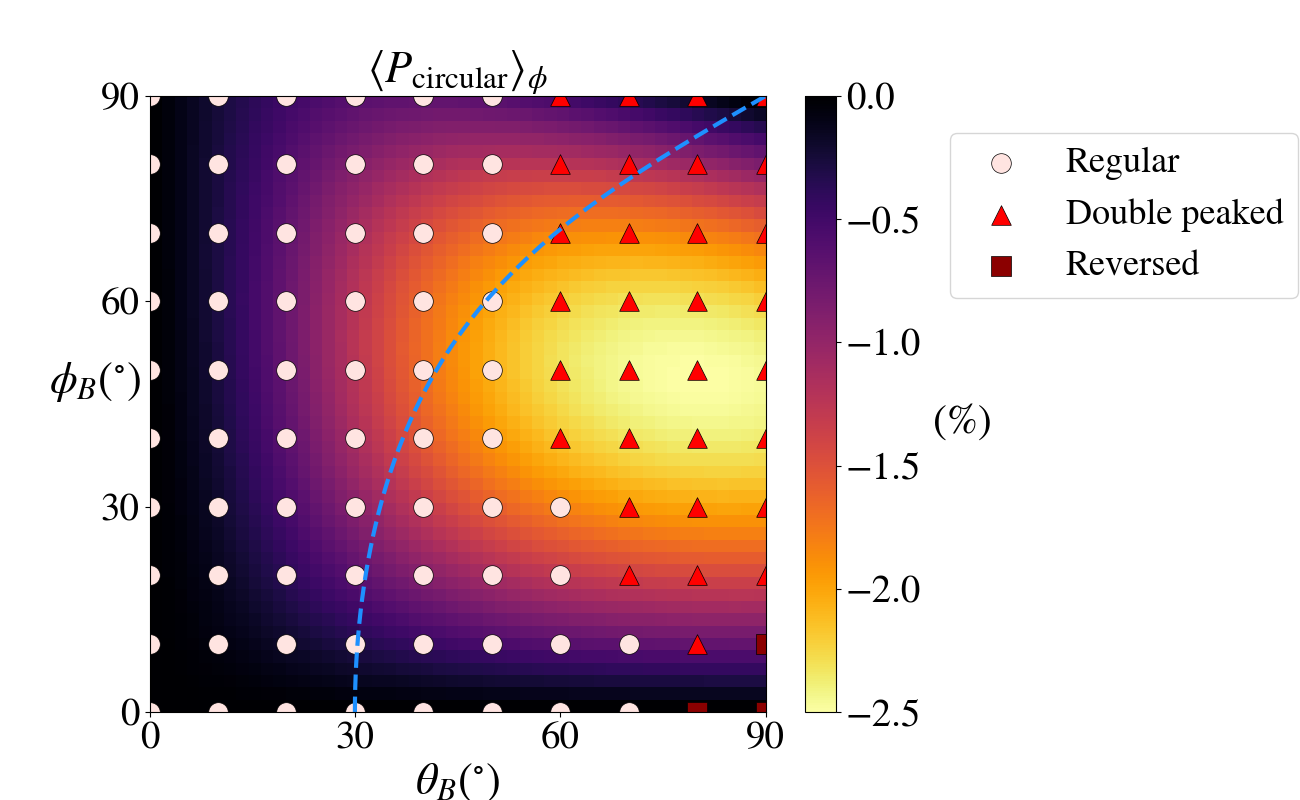}}
    \caption{Degree of circular polarization azimuthally averaged over a disk surface ring at $h/r=0.2$. We assumed $i=55^{\circ}$ and the mixture with amorphous carbon as dust composition. Each symbol represents a different category of azimuthal profile shape. The blue dashed line indicates the disk wind boundary: disk winds may occur at field configurations at the right side of the line.}
    \label{fig:heatmap_avg}
\end{figure}

To further analyze the disk CP patterns, we calculated the azimuthally averaged degree of circular polarization, given by 
\begin{equation}
    \left< P_{\textrm{circular}} \right> _{\phi} = \frac{1}{2\pi} \int_{0}^{2\pi} P_{\textrm{circular}}(\phi)d\phi,
\end{equation}
where $\phi$ is the disk azimuthal angle. We examined the integrated CP for a $50\times50$ grid of $\phi_{B}$ and $\theta_{B}$ values, assuming $i=55^{\circ}$ and the dust composition with amorphous carbon. Figure \ref{fig:heatmap_avg} shows the azimuthally averaged CP values for various magnetic field configurations. The majority of possible magnetic field angles produce an integrated negative degree of CP, although this could also be positive if we assume negative $\phi_B$ values. The averaged CP amplitude tends to become larger for larger $\theta_B$ at intermediate $\phi_B$ values. This suggests that even if the disk is spatially unresolved, CP signals can remain nonzero, particularly when the magnetic field makes a large angle with the $z$-axis. 

\begin{figure}
\resizebox{\hsize}{!}{\includegraphics{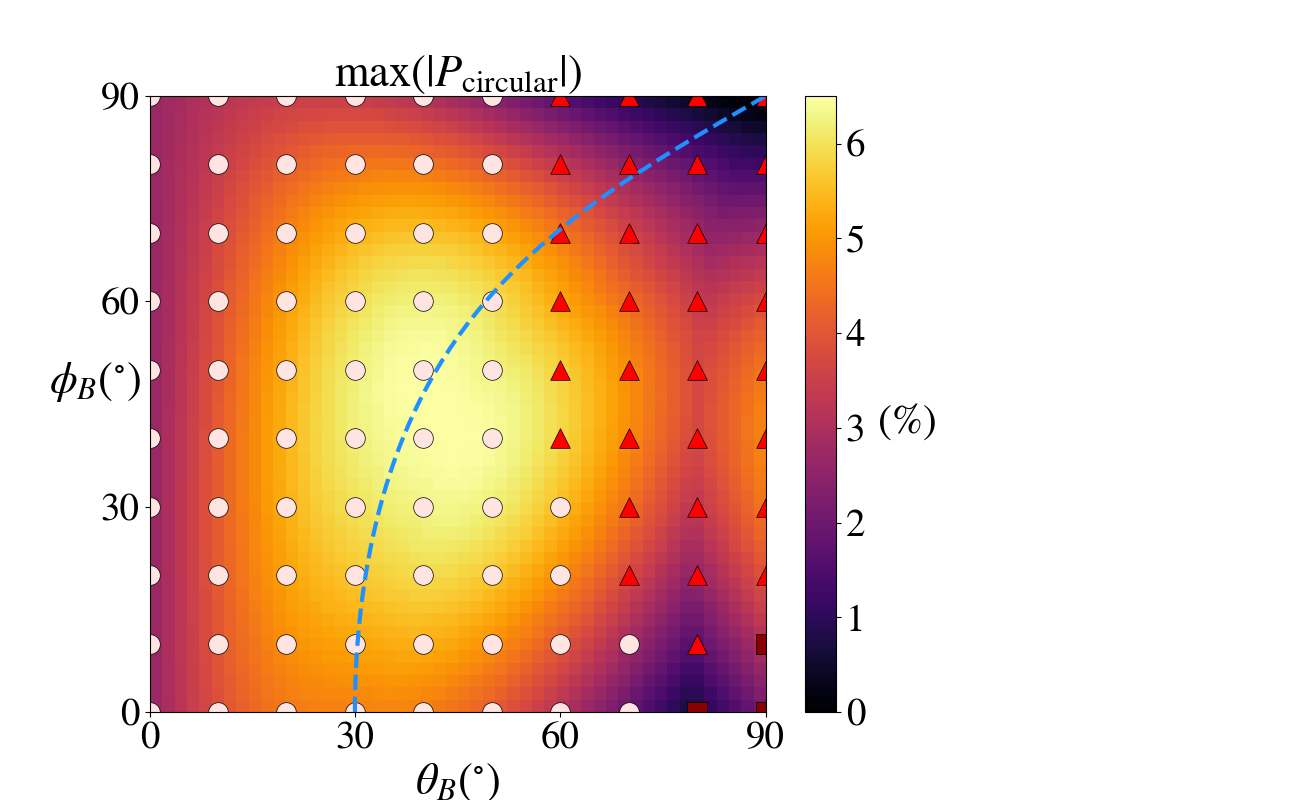}}
    \caption{Absolute value of the maximum degree of circular polarization of a disk surface ring at $h/r=0.2$. We assumed $i=55^{\circ}$ and the mixture with amorphous carbon as dust composition. Each symbol represents a different category of azimuthal profile shape. The blue dashed line indicates the disk wind boundary: disk winds may occur at field configurations at the right side of the line.}
    \label{fig:heatmap_max}
\end{figure}

The averaged CP percentage is zero in some situations: when the magnetic field is vertical ($\theta_B=0^\circ$), poloidal ($\phi_{B} = 0^{\circ}$), or toroidal ($\theta_B=90^\circ$, $\phi_B=90^\circ$). This can be caused by either an averaging effect or no intrinsic CP production at all. To distinguish which of these two causes is responsible, Fig. \ref{fig:heatmap_max} shows the nonaveraged maximum absolute value of CP at a disk surface ring of $h/r=0.2$. The nonzero maximum values for $\phi_{B}=0^{\circ}$ or $\theta_{B} = 0^{\circ}$ reveal that oblates aligned with these magnetic fields produce a symmetric CP pattern, where one side of the disk has negative CP values and the other side has positive (see Fig. \ref{fig:disks}, upper panel). As a result, CP signals cancel out when they are averaged. When the field is purely toroidal ($\theta_{B} = 90^\circ; \phi_{B} = 90^\circ$), the maximum CP is zero. Alignment with this magnetic field structure results in the symmetry axis of the oblate being perpendicular to the incident light, which results in no intrinsic circular polarization \citep{BandermannKemp1973}. Because of this, we conclude that the toroidal magnetic field seems to be better probed by the linear polarization technique instead of the CP-based technique (see Sect. \ref{sec:lin_circ}). In Fig. \ref{fig:heatmap_max}, we also find that there is another case where the intrinsic CP value becomes zero, that is, when $\theta_{B}=78^{\circ}$ and $\phi_{B}=0^{\circ}$. In this case, the incident light is parallel to the symmetry axis of the oblate grain. As a result, the $S_{41}$ component vanishes, and therefore no circular polarization is produced \citep{BandermannKemp1973}.

\subsection{Azimuthal profiles}
\begin{figure}
\resizebox{\hsize}{!}{\includegraphics{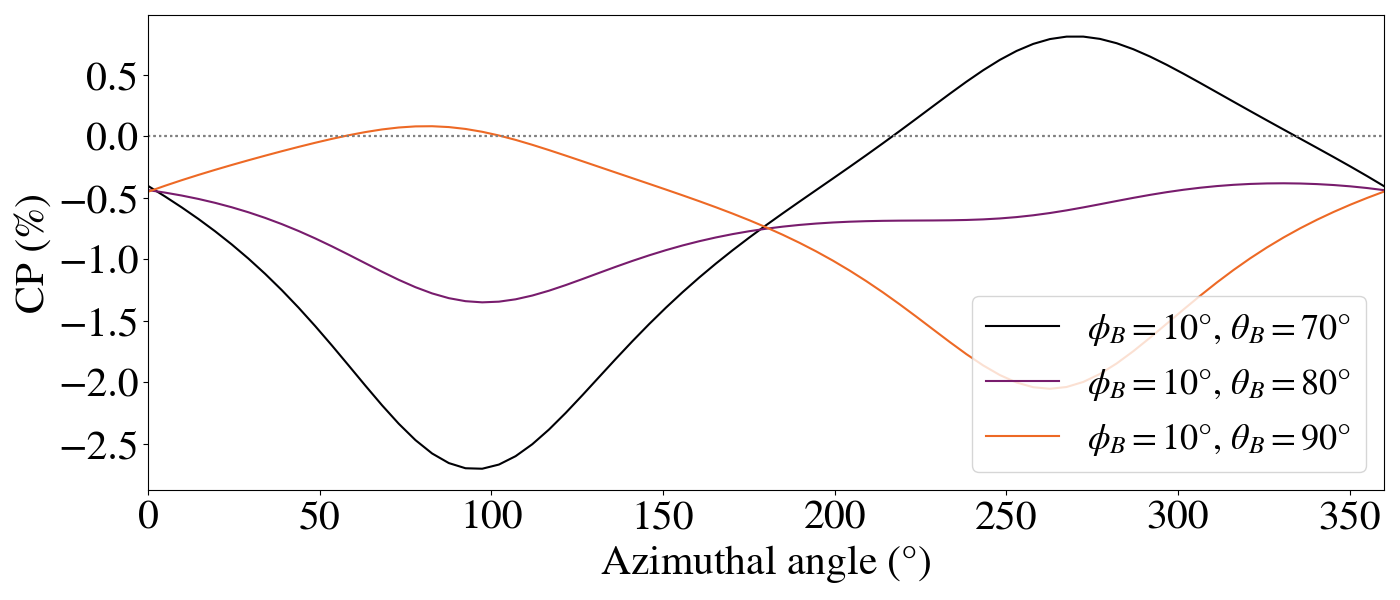}}
\caption{Azimuthal profiles of the degree of circular polarization produced by aligned grains at a disk surface ring with $i = 55^{\circ}$ and $h/r = 0.2$. The azimuthal angle is measured counterclockwise from the near side of the disk. The three shown $B$-field configurations each produce a different CP profile: `regular' (black), `reversed' (orange), and `double peaked' (purple).}
\label{fig:azimuth_profiles}
\end{figure}

To study the effect of the $B$-field on the CP pattern in more detail, we created azimuthal profiles. These profiles show the CP of a disk surface ring at $h/r = 0.2$, measured counterclockwise from the near side of the disk.
After evaluating all possible magnetic field angles, we distinguished three different shapes of azimuthal CP dependence: a dip followed by a peak (``regular''), a peak followed by a dip (``reversed''), and two dips (``double peaked''). We present examples for these three profile shapes in Fig. \ref{fig:azimuth_profiles}, where we assume the dust composition with amorphous carbon and an inclination of $i = 55^{\circ}$. Because we find that the reversed profile occurs in only a relatively small part of the parameter space, we mainly focus on the regular and doubled-peaked profiles hereafter.

We find that the regular and double-peaked azimuthal CP profiles can place additional constraints on the magnetic field configuration, as shown in Figs. \ref{fig:heatmap_avg} and \ref{fig:heatmap_max}. In those plots, we overlaid the distribution of the three azimuthal profile shapes. In addition, we indicated the boundary of magnetic field structures at which MHD disk winds by centrifugal force might occur with a blue dashed line. In order to launch a wind along the magnetic field lines, the centrifugal force needs to exceed the stellar gravity. This requirement is met when $\sin\theta_{B}\ge1/\sqrt{3\cos^{2}\phi_{B} + 1}$. When $\phi_{B}=0$, we obtain $\theta_{B}=30^{\circ}$, which agrees with the result of \cite{Blandford1982}. Magnetic field configurations at the right side of the line have field lines that make a sufficiently large angle with the $z$-axis to give rise to MHD disk winds by centrifugal force. Almost all double-peaked CP profiles are situated inside the disk-wind boundary, which suggests that these types of profiles are produced by magnetic field lines that make a large angle with the $z$-axis. 

\subsection{Effect of disk inclination} 
\begin{figure}
\resizebox{1.2\hsize}{!}{\includegraphics{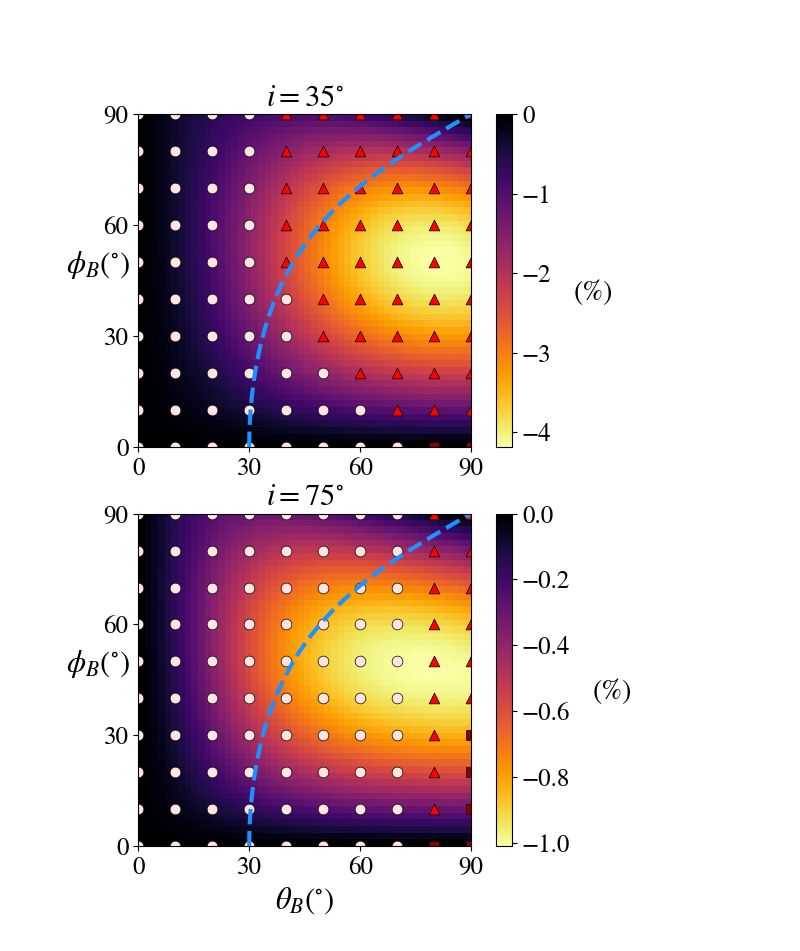}}
    \caption{Same as Fig. \ref{fig:heatmap_avg}, but we assumed different inclinations: $i=35^{\circ}$ for the upper panel and $i=75^{\circ}$ for the lower panel. The boundary between the regular and double-peaked profiles shifts to the left/right for lower/higher inclinations.}
    \label{fig:inclination}
\end{figure}

In order to study the impact of disk inclination angle, we also compare CP at $i=35^\circ$ and $75^\circ$ in Fig. \ref{fig:inclination}. The effect of disk inclination can be seen in the differences between the boundaries separating the regular and double-peaked profiles. For the case of $i=35^\circ$, the double-peaked profile occurs even if $\theta_{B}$ is as small as $40^\circ$. For the case of $i=75^\circ$, the double-peaked profiles only occur at even higher $\theta_B$ values than in a lower inclination case. Therefore, as long as the disk is moderately or highly inclined, the occurrence of the double-peaked profile indicates a magnetic field configuration with a high $\theta_B$ value that is favorable for launching winds by centrifugal force.

\subsection{Effect of other dust-grain parameters} \label{sec:other}
We examine how different dust models would affect our results by evaluating the CP patterns for another dust composition and other axis ratios. In the alternative dust composition, we replaced amorphous carbon with refractory organics \citep{Birnstiel2018}. We find that this less absorbing material results in less circular polarized light, with a maximum value of $\sim\mkern-4mu 0.4\%$ compared to $\sim\mkern-4mu 6\%$ for the mixture with amorphous carbon. These lower CP values can be physically explained as follows. Circular polarization is produced when there is a phase lag between dipole radiation along the major and minor axis of the grain. The amount of the phase lag is related to the absorption efficiency and becomes zero when grains are nonabsorbing \citep{BandermannKemp1973}. Consequently, less absorbing material results in a smaller phase lag and therefore weaker CP production. Lower axis ratios also result in less CP because circular polarization arises from the difference in polarizability along each spheroid axis, which is smaller for more spherical particles. Fortunately, although both grain composition and axis ratio affect the absolute value of CP, the shape of the azimuthal profiles does not change. This demonstrates that the occurrence of double-peaked profiles is independent of the dust composition and grain-axis ratio, and is therefore a reliable probe of the magnetic field structure, as long as the Rayleigh approximation holds. Therefore, if we detect CP, its amplitude would be a useful diagnostic of grain properties. 

As long as we work on the Rayleigh limit, the effects of axis ratio and dust composition on the CP amplitude are degenerate. To better constrain dust properties, a combined analysis with the degree of CP and disk-scattered flux would be useful. This is because the imaginary part of the refractive index is not only important for CP, but also for the single scattering albedo, which in turn strongly affects disk-scattered flux.

\section{Discussion} \label{sec:discussion}

\subsection{Role of linear and circular polarization in probing disk magnetic fields} \label{sec:lin_circ}
Our study, together with previous studies, demonstrates that polarization signals in scattered optical and NIR light contain valuable information about the $B$-field in protoplanetary disks. \cite{yangli2022} explored how linearly polarized scattered light can be used to diagnose the magnetic field. Here, we discuss how two types of polarization observations (linear and circular) can complement each other and advance our understanding of disk magnetic fields. 

It would be very useful to combine the linear and circular polarization analysis because this allows us to exclude polarization signals that are produced via pathways other than magnetically aligned grains, such as multiple scattering \citep{Canovas2015} or instrumental errors. Observing both linear and circular polarization provides an independent check as to whether or not the polarization signals are truly produced by aligned grains. 

In addition, linear and circular polarization in  disk-scattered NIR light are sensitive to different magnetic field components. Linear polarization angles seem more suitable to indicate the presence of the toroidal component, whereas circular polarization profiles better probe the poloidal component with a large angle to the $z$-axis. The former $B$-field configuration drives the wind by a magnetic pressure gradient \citep[e.g.,][]{Bai2017}, while the latter does this via centrifugal force \citep{Blandford1982}. Therefore, the combination of these could narrow down the possible magnetic field structures and eventually reveal a launching mechanism of MHD winds.

To study the linear and circular polarization signals as described in the techniques of our study and of \cite{yangli2022}, the disk needs to be resolved. However, the azimuthally averaged CP values can be measured without resolving the disk. We find that an integrated nonzero degree of circular polarization already points towards the presence of aligned grains at the disk surface, which is an essential requirement to study the magnetic field structure using linear and circular polarization. The integrated CP value can reveal disks that meet this requirement, which helps to search for suitable targets.

\subsection{Circular polarization from large aligned grains} \label{sec:large}
In this study, we adopt the Rayleigh approximation, which is only applicable for grains smaller than the wavelength. At a wavelength of $\lambda=1.6~\mu$m, the Rayleigh approximation is applicable when the grain radius is smaller than $\lambda/2\pi\sim0.25~\mu$m. Here, we discuss how the presence of larger grains would affect the CP and whether or not the Rayleigh limit is a valid assumption. 

\cite{Fukushima2020} calculated circularly polarized scattered light by aligned grains in interstellar slab-like clouds. These authors found that increasing grain radius results in increasing the degree of CP. They also found that the CP pattern outside the Rayleigh limit would become less simple than in the Rayleigh limit (see their Fig. 4). Therefore, the presence of micron-sized grains at the disk surface would make much more complicated CP patterns  than those presented in this study.

However, one caveat of \cite{Fukushima2020} is that the authors only considered compact micron-sized grains. At the disk surface, micron-sized dust particles at the disk surface would be aggregates \citep[e.g.,][]{DominikTielens1997,BlumWurm2008}, which are clusters of small single particles (monomers) glued together by intermolecular forces, such as the van der Waals force. Although these studies suggest that aggregates may have sizes beyond the Rayleigh limit, the degree of polarization of scattered light by large aggregates rather resembles that of individual monomers \citep[e.g.,][]{West1991,Kozasa1993,Volten2007,Tazaki2022}. \cite{Tazaki2022} estimated the upper limit on the monomer radius to be $0.4 \, \rm{\mu m} $ for a set of protoplanetary disks based on polarimetric observations. Therefore, as long as such dust aggregates are the dominant population of the disk surface, and monomers in the aggregate are aligned, our argument would remain approximately valid. Another way to circumvent possible deviations from the Rayleigh solution would be to observe the disk at longer wavelengths, such as $2.2 \, \mu$m ($K$-band). 

\subsection{Do grains align with magnetic fields?}
In the above sections, we assume that grains at the disk surface perfectly align with magnetic fields. Here, we discuss to what extent this assumption is reasonable. 

\cite{tazaki2017} and \cite{yangli2022} suggest that grain alignment with the magnetic field, also known as $B$-RAT, is feasible at the disk surfaces when grains contain a sufficient amount of superparamagnetic inclusions. However, $B$-RAT is not the only known mechanism to align grains. Dust grains at the disk surface can also become aligned with the stellar radiation, which is called $k$-RAT \citep{LazarianHoang2007}. $k$-RAT alignment predicts that the grain's minor axis becomes parallel to the incoming radiation direction. In this case, we expect no circular polarization as long as the grain polarization property is approximated by the Rayleigh limit. This is because no CP is produced when the particle symmetry axis is either in the scattering plane ($k$-RAT for oblate grains) or perpendicular to the incident light direction ($k$-RAT for prolate grains) \citep{BandermannKemp1973}.

Therefore, if CP observations end up with a nondetection, this could be attributed to $k$-RAT alignment, although other reasons still exist, such as very low alignment efficiency, $B$-RAT alignment with a symmetric $B$-field structure, and so on. Conversely, if we detect a CP signal, this is likely arising from $B$-RAT alignment. In this way, CP observations will shed light on grain-alignment processes at the disk surface.

It is also worth mentioning that CP can be produced by multiple scattering from nonaligned or spherical grains. However, such a CP signal will efficiently vanish when it is integrated over the disk as long as we can assume an axisymmetric disk structure. In contrast, as we discuss in Sect. \ref{sec:integrated}, an integrated CP signal can be nonzero if it is due to aligned grains. Therefore, if we detect a net nonzero integrated CP from an axisymmetric disk, it is likely attributable to grains aligned with the $B$-field rather than the multiple-scattering origin.

\subsection{Future prospects}
The next step in this study would be to compare our results with direct observations of the degree of circular polarization in protoplanetary disks. No direct circular polarization measurements of protoplanetary disks have been published yet, but promising instruments and techniques are emerging. The Subaru Coronagraphic Extreme Adaptive Optics (SCExAO) instrument on the Subaru telescope will allow us to obtain circular polarization images of resolved protoplanetary disks at NIR wavelengths. In addition, although not originally designed for circular polarization measurements, VLT/SPHERE could also be used to observe CP, as proposed by \cite{Holstein2020}.

One caveat of our model is that we assume single scattering to simplify the disk radiative transfer processes. However, protoplanetary disks are optically thick, and in general, multiple scattering plays a role, which can affect the degree of polarization \citep[e.g.,][]{MaSchmid2022}. In addition, aligned grains also produce nonazimuthal linear polarization \citep{yangli2022} which can be converted into circular polarization via multiple scattering. Further clarification of these effects is left for our future study.

\section{Summary} \label{sec:summary}
In this study, we explored a new method to study the magnetic field structure at the surface of protoplanetary disks using NIR circular polarization of scattered light. Assuming a disk model and various magnetic field geometries, we created synthetic maps of the degree of circular polarization produced by scattering oblate spheroids at the disk surface. By analyzing the azimuthal dependence and the integrated value, we studied to what extent the configuration of the $B$-field is reflected in the CP patterns. We also investigated the effects of disk inclination and dust grain parameters on our results. We examined what we can learn from the CP disk images about the launching mechanism of MHD disk winds, and how this method ties in with linear polarization techniques. Our primary findings are as follows.  
\begin{itemize}
\item It is shown that magnetically aligned grains at the disk surface can produce circular polarization signals. Also, circular polarization signals remain nonzero even if the disk is not spatially resolved. 
\item It is found that a disk exhibiting a CP pattern with a double-peaked azimuthal profile likely has a poloidal magnetic field with a large angle with respect to the $z$-axis. In contrast, circular polarization becomes zero for a toroidal configuration.
\item The degree of CP strongly depends on dust composition. A more absorbing composition, such as when grains are comprised of amorphous carbon, produces a higher degree of CP. The composition does not change the CP azimuthal profiles, as long as the Rayleigh approximation holds.
\item We find that circular polarization is complementary to the linear polarization technique to probe the disk $B$-field because the combined analysis allows us to break degeneracy, and the polarization signals trace different magnetic field components.

\end{itemize}
 
\begin{acknowledgements}
We thank an anonymous referee for useful comments. In addition, we thank C. Dominik and J. Hashimoto for a fruitful discussion. R.T. acknowledges the JSPS overseas research fellowship.
\end{acknowledgements}

%
  \bibliographystyle{aa} 
  \bibliography{refs.bib} 
%

\end{document}